\documentclass[lettersize,journal]{IEEEtran}
\usepackage{amsmath,amsfonts}
\usepackage{algorithmic}
\usepackage{algorithm}
\usepackage{array}
\usepackage[caption=false,font=normalsize,labelfont=sf,textfont=sf]{subfig}
\usepackage{textcomp}
\usepackage{stfloats}
\usepackage{url}
\usepackage{verbatim}
\usepackage{graphicx}
\usepackage{cite}
\usepackage{bm}

\usepackage{flushend,cuted}

\allowdisplaybreaks[4]

\begin{document}

\title{Joint Computing Offloading and Resource Allocation for Classification Intelligent Tasks in MEC Systems}

\author{Yuanpeng Zheng,~\IEEEmembership{Student Member,~IEEE,} Tiankui Zhang,~\IEEEmembership{Senior Member,~IEEE,} Jonathan Loo, Yapeng Wang, and Arumugam Nallanathan,~\IEEEmembership{Fellow,~IEEE}
\thanks{Part of this work has been presented at the IEEE Wireless Communications and Networking Conference (WCNC), Glasgow, Scotland, UK, Mar., 2023 \cite{ref32}.}
\thanks{This work is supported by Beijing Natural Science Foundation under Grants 4222010.
(Corresponding author: Tiankui Zhang)}
\thanks{Yuanpeng Zheng, Tiankui Zhang are with the
School of Information and Communication Engineering,
Beijing University of Posts and Telecommunications, Beijing 100876, China (e-mail: \{zhengyuanpeng, zhangtiankui\}@bupt.edu.cn).}
\thanks{Jonathan Loo is with the School of Computing and Engineering, University of West London,  London W5 5RF, U.K. (e-mail: jonathan.loo@uwl.ac.uk).}
\thanks{Yapeng Wang is with Faculty of Applied Sciences, Macao Polytechnic University, Macao SAR, China (e-mail: yapengwang@mpu.edu.mo).}
\thanks{Arumugam Nallanathan is with the School of Electronic Engineering and Computer Science, Queen Mary University of London, London E1 4NS, U.K. (e-mail: a.nallanathan@qmul.ac.uk).}  

}

\maketitle

\begin{abstract}
Mobile edge computing (MEC) enables low-latency and high-bandwidth applications by bringing computation and data storage closer to end-users. 
Intelligent computing is an important application of MEC, where computing resources are used to solve intelligent task-related problems based on task requirements. 
However, efficiently offloading computing and allocating resources for intelligent tasks in MEC systems is a challenging problem due to complex interactions between task requirements and MEC resources.
To address this challenge, we investigate joint computing offloading and resource allocation for intelligent tasks in MEC systems. 
Our goal is to optimize system utility by jointly considering computing accuracy and task delay to achieve maximum system performance. 
We focus on classification intelligence tasks and formulate an optimization problem that considers both the accuracy requirements of tasks and the parallel computing capabilities of MEC systems. 
To solve the optimization problem, we decompose it into three subproblems: subcarrier allocation, computing capacity allocation, and compression offloading. 
We use convex optimization and successive convex approximation to derive closed-form expressions for the subcarrier allocation, offloading decisions, computing capacity, and compressed ratio.
Based on our solutions, we design an efficient computing offloading and resource allocation algorithm for intelligent tasks in MEC systems.
Our simulation results demonstrate that our proposed algorithm significantly improves the performance of intelligent tasks in MEC systems and achieves a flexible trade-off between system revenue and cost considering intelligent tasks compared with the benchmarks.

\end{abstract}

\begin{IEEEkeywords}
Computing offloading, intelligent tasks, mobile edge computing, resource allocation.
\end{IEEEkeywords}

\section{Introduction}
With the rapid development of mobile edge computing (MEC), which supports not only communication but also computing and storage, more and more new applications such as computer vision, natural language processing, semantic communication, etc., are emerging constantly. 
By being closer to users of network than traditional cloud computing, MEC can obviously reduce application completion time and improve the quality of user experience with specific tasks.
In the context of the increasing number of intelligent computing scenarios for intelligent tasks, MEC needs to tackle with many related problems with specific characteristics \cite{ref1} which is diverse and different from traditional resource allocation problems.
However, few existing works consider the various requirements of those characteristics such as computing accuracy \cite{ref2} and parallel computing \cite{ref3} while considering resource allocation problems in MEC. 
Obviously, how to efficiently allocate diversified resource to support the specific demands of intelligent tasks is still an unaddressed problem.

Hence, in the context of massive Internet of Things (IoT) devices deployment and limited terminal computing capacity, characteristics of computing tasks are increasingly complex and their impact escalates in MEC. Existing works on computing offloading and resource allocation in MEC systems has become specific and multidimensional\cite{ref4,ref5,ref6,ref10,ref8,ref9,ref7}. 
C. Wang \textit{et al.}\cite{ref4} considered computation offloading and content caching strategies in wireless cellular network with MEC and formulate the total revenue of the network. 
With considering computing offloading and large data volume, Y. Ding \textit{et al.}\cite{ref5} propose a novel online edge learning offloading scheme for UAV-assisted MEC secure communications, which can improve the secure computation performance.
Considering edge cache-enabling UAV-assisted cellular networks, T. Zhang \textit{et al.}\cite{ref6} formulated a joint optimization problem of UAV deployment, caching placement and user association for maximizing quality of experience of users, which is evaluated by mean opinion score.
J. Feng \textit{et al.}\cite{ref10} considered the stochastic nature of tasks and proposed a framework that maximizes revenue of network providers in MEC systems, where a multi-time scale scheme was adopted to increase revenue on the basis of QoS guarantee.
As an important scene, J. Y. Hwang \textit{et al.}\cite{ref8} introduced the IoT platform with an efficient method of integrating MEC and network slice to maximize the effect of decreasing delay and traffic prioritization. 
W. Lu \textit{et al.}\cite{ref9} proposed two secure transmission methods for multi-UAV-assisted mobile edge computing based on the single-agent and multi-agent reinforcement learning respectively and achieve larger system utility. 
T. Zhang \textit{et al.}\cite{ref7} proposed a new optimization problem formulation that aims to minimize the total energy consumption including communication-related energy, computation-related energy and UAV's flight energy by optimizing the bits allocation, time slot scheduling, and power allocation as well as UAV trajectory design with multiple computation strategies.
As shown above, these works studied characteristics of tasks in MEC systems under the assumption of simplified computing process. Nevertheless, in actual application scenarios, intelligent tasks have complicated characteristics and demands including complexity, accuracy and parallelism that will cause many changes, which need to be considered in the computing offloading and resource allocation in MEC systems.

With the increasing trend of artificial intelligence in recent years, as a form of computing that solves practical problems by optimizing existing computing methods and resources systematically and holistically according to task requirements\cite{ref31}, intelligent computing has brought more requirements to the MEC field. The researches on intelligent computing for intelligent tasks in MEC are getting more attention\cite{ref1,ref15,ref16,ref17,ref18}. 
H. Xie \textit{et al.}\cite{ref1} investigated the deployment of semantic communication system based on edge and IoT devices where MEC servers computing the semantic model and IoT devices collect and transmit data based on semantic task model. 
X. Ran \textit{et al.}\cite{ref15} focused on a framework that ties together front-end devices with more powerful backend helpers to allow intelligent tasks to be executed locally or remotely in the edge with considering the complex interaction between computing accuracy, video quality, battery constraints, network data usage and network conditions.
By considering the computer vision to video on cloud-backed mobile devices using intelligent tasks, S. Han \textit{et al.}\cite{ref16} studied the resource management including strain device battery, wireless data and cloud cost budgets in MEC systems.
M. Jankowski \textit{et al.}\cite{ref17} introduced the intelligent tasks of image retrieval problem at the wireless edge which maximizes the accuracy of the retrieval task under power and bandwidth constraints over the wireless link.
X. Xu \textit{et al.}\cite{ref18} indicated there are increasing gaps between the computational complexity and energy efficiency required by data scientists and the hardware capacity made available by hardware architects while processing intelligent tasks in edge and discussed various methods that help to bridge the gaps.
Apparently, the research of intelligent computing for specific intelligent tasks in MEC has become extensive and gradually mature, which represents resource allocation for intelligent tasks in MEC has a certain basis.

In this context, some studies considering the complicated characteristics of specific intelligent tasks above become important and lay the foundation for resource allocation in MEC\cite{ref2, ref3, ref11, ref12, ref13, ref14}.
B. Gu \textit{et al.}\cite{ref2} investigated the fitting of modelling classification accuracy by verification of large data sets for intelligent tasks and find that power law is the best among all models.
As the computing tasks become complicated gradually, D. Bruneo \textit{et al.}\cite{ref3} studied the requirements of computing infrastructures and introduced the parallel computing model for complex computing tasks with the corresponding quality of service.
Considering the scenario of intelligent tasks, H. Xie \textit{et al.}\cite{ref11} proposed a brand new framework of semantic communication where a deep learning based system for text transmission combined with natural language processing and semantic layer communication was constructed.
Analogously, M. Bianchini \textit{et al.}\cite{ref12} and D. Justus\textit{et al.}\cite{ref13} indicated that the prediction of execution time of intelligent tasks depends on many influence factors including the construction of tasks and hardware features. 
In particular, W. Fan \textit{et al.}\cite{ref14} consider a quality-aware edge-assisted machine learning task inference scenario and propose a resource allocation scheme to minimize the total task processing delay considering the inference accuracy requirements of all the tasks.
Obviously, in the face of specific intelligent tasks such as DNN training and inference, MEC can take advantage of control and optimization to perform more tasks at low cost while considering intelligent computing scenarios.

\subsection{Motivation and Contribution}
As mentioned above, the combination of computing offloading and resource allocation in MEC systems considering the demands of intelligent tasks is still an unaddressed research area, which motivates this contribution. 
In this paper, considering the intelligent computing for intelligent tasks, we study specific classification intelligence tasks and adopt the training accuracy fitting model\cite{ref2} and parallel computing model of hardware servers\cite{ref3} as key influencing factors. 
Our scenario is based on classification intelligent tasks in \cite{ref2} considering lightweight distributed machine learning training and parallel computing in MEC systems combined with control and optimization through communication.
We model the key indicator, i.e., computing accuracy into our resource allocation algorithm and make the trade-off with task delay, which improves the applicability of our model in intelligent task scenarios.
Though we focus on specific intelligent tasks due to the differences between tasks, our method is also proper for other similar applications.
The main contributions of this paper are as follows:
\begin{itemize}
  \item{We formulate an optimization problem that considers both the accuracy requirements of tasks and the parallel computing capabilities of MEC systems. We adopt the training accuracy fitting model of classification intelligence tasks and parallel computing model of hardware servers to represent the precise demands while considering computing offloading and resource allocation. We define the system utility which consists of the system revenue depending on computing accuracy and cost depending on task delay. Therefore, an integrated framework for computing offloading and resource allocation for intelligent tasks has the potential to improve the intelligent computing performance of the MEC systems.}
  \item{We solve the highly coupled computing offloading and resource allocation problem by decoupling manifold optimization variables through the idea of iterative optimization into three subproblems: subcarrier allocation, computing capacity allocation and compression offloading. We design an efficient computing offloading and resource allocation algorithm for intelligent tasks in MEC systems where the subcarrier allocation problem and compression offloading problem are solved by successive optimization approximation and the computing capacity allocation problem is solved by convex optimization. We derive closed-form expressions for all variables and acquire the suboptimal solution through iterative optimization.}
  \item{We demonstrate the simulation results which verify that our framework is applicable to intelligent computing scenario in MEC systems. It is shown that the proposed algorithm significantly improves the performance of intelligent tasks in MEC systems and achieves a flexible trade-off between system revenue and cost considering intelligent tasks compared with the benchmarks.}
\end{itemize}

\subsection{Organization}
The rest of this paper is organized as follows. In Section II, the system model and utility function are introduced. 
In Section III, the problem formulation and decomposition into several subproblems of the problem are represented to design our algorithm. 
The performance of the proposed algorithm is evaluated by the simulation in Section IV, which is followed by the conclusions in Section V.

\section{System Model}
We consider the heterogeneous cellular scenario as shown in Fig. 1(a) and equip MEC servers on small base stations (SBS) to form the MEC systems. We set the total amount of users is $U$. The set of MEC systems is denoted by $K^S = \{1,...,k,...,K\}$ and it is assumed that SBS $k$ is associated with $U_k$ mobile users. We let $U^S_k = \{1,...,u_k,...,U_k\}$ denote the set of users associating with SBS $k$ where $u_k$ refers to the $u$th user which associates with the $k$th SBS. 
The set of classification intelligent tasks is denoted by $M^S = \{1,...,m,...,M\}$ which has the property of parallel computing, i.e., task $m$ requires parallel processing on the server, and the requirement of accuracy. 
In our model, we consider two types of computing including local computing and offloading to MEC computing in our systems as shown in Fig. 1(b). We adopt few-shot learning and data compression in step $1$ in Fig. 1(b) which is applied to size compression in traditional classification intelligent tasks and feature extraction in semantic tasks for training. For example, the image set collected by the local device needs to be classified and recognized, therefore local intelligent device will choose local computing or offloading to edge computing where image set needs size compression or semantic feature extraction in our model.
Let the bandwidth resource of our system be $B$, computing capacity of local device be $F^L_{u_k}$, computing capacity of the MEC server be $F_k$ and delay limit for computing task $m$ be $\tilde{t}_m$.

\begin{figure*}[!t]
  \centering
  \subfloat[]{\includegraphics[scale=0.75]{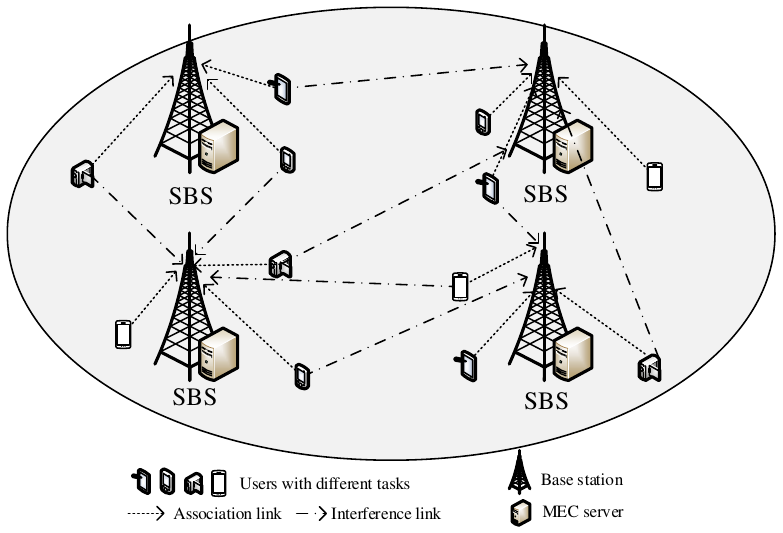}%
  \label{fig_1a}}
  \hfil
  \subfloat[]{\includegraphics[scale=0.4]{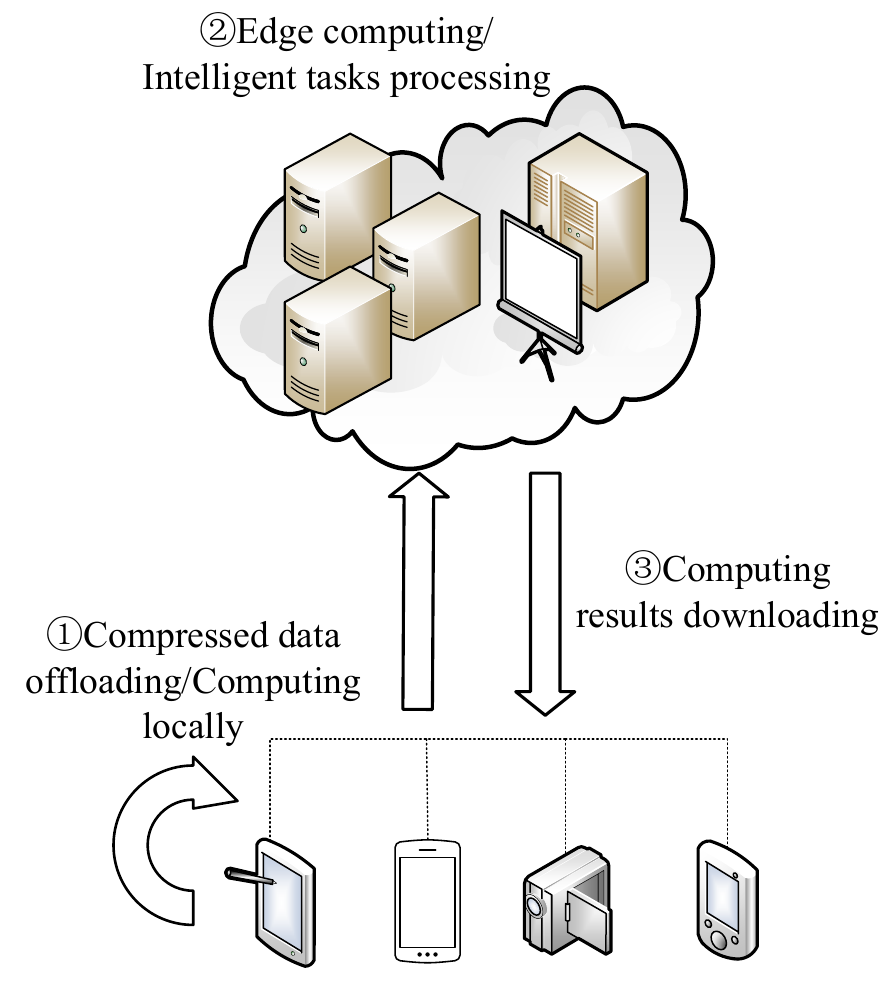}%
  \label{fig_1b}}
  \caption{The system model. (a) The system scenario. (b) Computing task processing flow of the MEC server.}
  \label{fig_1}
\end{figure*}

\begin{table}[h]
  \renewcommand\arraystretch{1.5}
  \caption{Main Symbol and Variable List}
  \centering
  \begin{tabular}{|p{1.2cm}|p{5.5cm}|}
  \hline
  {\textbf{Notation}} & \textbf{Description} \\
  \hline
  $U$ & Total Number of users  \\
  \hline
  $K$ & Number of SBSs  \\
  \hline
  $U_k$ & Number of users associating with SBS $k$ \\
  \hline
  $M$ & Number of tasks  \\
  \hline
  $N$ & Number of subcarriers \\
  \hline
  $B$ & Total bandwidth resource  \\
  \hline
  $F_k$ & Computing capacity of each MEC  \\
  \hline
  $F^L_{u_k}$ & Computing capacity of local device  \\
  \hline
  $\tilde{t}_m$ & Delay limit for computing tasks \\
  \hline
  $\tilde{y}_m$ & Limit of computing accuracy of computing tasks  \\
  \hline
  $x_{u_k}$ & Indicator of whether user $u_k$ offloading its tasks \\
  \hline
  $z_{u_km}$ & Indicator of whether task $m$ requested by user $u_k$ \\
  \hline
  $R$ & The system utility \\
  \hline 
  $\rho_{u_kn}$ & Indicator of whether subcarrier $n$ allocated to user $u_k$\\
  \hline
  $r_{u_k}$ & Uplink transmission rate of user $u_k$\\
  \hline 
  $f^O_{uk}$ & The computing capacity allocated to user $u_k$ by MEC servers\\
  \hline
  $t_{u_k}^{comm}$ & Transmission delay of the data compressed by user $u_k$ in the wireless link\\
  \hline
  $T^L_{u_k}$ & Local computing delay of user $u_k$\\
  \hline
  $t_{u_k}^{comp}$ & The computing latency incurred by user $u_k$ to perform tasks on MEC servers\\
  \hline
  $t_{u_k}$ & Total task delay of user $u_k$\\
  \hline
  $a_{u_k}$ & The raw data collected by user $u_k$\\
  \hline
  $b_{u_K}$ & Compressed data of user $u_k$\\
  \hline
  $\varepsilon_{u_k}$ & Compression ratio of user $u_k$\\
  \hline
  $y(\cdot )$ & Computing accuracy of corresponding tasks\\
  \hline
  $L$ & Weight parameter between system revenue and cost\\
  \hline
\end{tabular}
\end{table}

\subsection{Communication Model}
In our system, every SBS in the network is equipped with the MEC server, so each user can offload its computing task to the MEC server through the SBS to which it is connected. We denote $x_{u_k} \in \{0,1\}, \forall u,k$ as the computing offloading indicator variable of user $u_k$. Specially, $x_{u_k} = 1$ if user $u_k$ offload its computing task to the MEC server via wireless network and we have $x_{u_k} = 0 $ if user $u_k$ determine to compute its task locally on the mobile device. Therefore, we denote $\bm{x} = \{x_{u_k}\}_{u_k \in U^S_k, k \in K^S}$ as the offloading indicator vector.

In this paper, we consider that spectrum used by SBSs is overlaid and spectrum within one SBS is orthogonally assigned to every user. Therefore, there exists interference between SBSs but there will be no interference within one SBS. We only analyse uplink transmission and divide the total spectrum into $N$ subcarriers, which is denoted as $N^S = \{1,...,n,...,N\}$. All users can reuse subcarriers for uplink transmission to improve spectrum utilization. We denote $\rho_{u_kn}\in \{0,1\},\forall u,k,n$ as subcarrier allocation variables, where $\rho_{u_kn} = 1$ means subcarrier $n$ is allocated to user $u_k$ which is associated with SBS $k$ and $\rho_{u_kn} = 0$ otherwise. Obviously, one subcarrier on an SBS can only be allocated to one user at a time, therefore we have $\sum^{U_k}_{u_k=1}\rho_{u_kn} \leq 1,\forall n$. We denote $\bm{\rho} = \{\rho_{u_kn} \}_{{u_k}\in U^S_k, k\in K^S,n\in N^S}$ as the subcarrier allocation vector. Then, the uplink transmission rate of user $u_k$ on subcarrier $n$  given as
\begin{equation}
  \label{deqn_comm}
  r_{u_kn} = \frac{B}{N}log_2\left( 1 + \frac{P_{u_kn}g_{u_kn}}{I_{u_kn} + \sigma^2} \right), \forall u,k,n,
\end{equation}
where $P_{u_kn}$ represents transmit power from user $u_k$ to SBS $k$, $g_{u_kn}$ represents wireless channel gain between user $u_k$ and SBS $k$ on subcarrier $n$, and $I_{u_kn}$ represents co-channel interference of users associating with other SBSs on the same frequency of user $u_k$, which is given as $I_{u_kn} = \sum_{c\in K^S,c\neq k} \sum^{U_c}_{u'_c=1} \rho_{u'_cn}g_{u'_cn}P_{u'_cn},\forall u,k,n$. $\sigma^2$ denotes the noise power of additive white Gaussian noise. Obviously, the uplink transmission rate of user $u_k$ is denoted as $r_{u_k} = \sum^N_{n=1} \rho_{u_kn}r_{u_kn},\forall u,k$, which means the total transmission rate of user $u_k$ is the summation of the transmission rate on all subcarriers which are allocated to user $u_k$. 

\subsection{Computing Model}
For the computing model, we consider each user $u_k$ has a computing task $m$, and denote $z_{u_km} \in \{0,1\}, \forall u,k,m$ as the indicator variable of the computing task $m$ of user $u_k$. Specially, $z_{u_km} = 1$ if the computing task of user $u_k$ is $m$, otherwise $z_{u_km} = 0$. 
In our model, we assume that $z_{u_km}$ is already given as the user request and we have $\sum^M_{m=1} z_{u_km} = 1,\forall u,k$. 
Obviously we consider that one task can be requested by each user at a time, which make our model clearer for task properties of accuracy requirement and parallel computing. The scenario of multi-task case can be acquired after small modified base on our model.
We consider two types of computing approaches, i.e., local computing and task offloading.

\emph{1) Local Computing:} For the local computing approach, the raw data of user $u_k$ is given as $a_{u_k}$, and we can acquire the computing delay through the raw data $a_{u_k}$ directly, which is given as
\begin{equation}
  \label{deqn_lcd}
  T^L_{u_k} = \frac{\sum \limits ^M_{m=1}z_{u_km}F_{u_km}\left(a_{u_k}\right)}{F^L_{u_k}}, \forall u,k,
\end{equation}
where $F_{u_km}(\cdot )$ represent the computing resource overhead of corresponding data volume after parallel computing. In this paper we think of it as approximate linear relationship after the property of parallel computing of tasks has been considered.
Note that we neglect power consumption of computing in our model, therefore we do not consider extra computing cost in local device.

\emph{2) Task Offloading:} For the task offloading approach, user $u_k$ will compress the raw data $a_{u_k}$ to 
\begin{equation}
  \label{deqn_cd}
  b_{u_k} = \frac{a_{u_k}}{\varepsilon _{u_k}},\forall u,k,
\end{equation}
where $\varepsilon _{u_k}$ is denoted as compression ratio of user $u_k$ and $\varepsilon _{u_k}\geq 1, \forall u,k$. We denote $\bm{\varepsilon} = \{ \varepsilon_{u_k}\}_{u_k\in U^S_k, k\in K^S}$ as the compression ration vector. Apparently, we have $\alpha_{u_k} = (1-x_{u_k})a_{u_k} + x_{u_k}b_{u_k}$. Then, the compressed data $b_{u_k}$ is transmitted to SBS $k$ to process and compute, and the transmission delay of the compressed data from user $u_k$ in wireless link is given as
\begin{equation}
  \label{deqn_trd}
  t^{comm}_{u_k} = \frac{b_{u_k}}{r_{u_k}},\forall u,k.
\end{equation}
Let $f^O_{u_k}$ be computing capacity allocated to user $u_k$ from SBS $k$ and $\bm{f^O} = \{f^O_{u_k}\}_{u_k\in U^S_k, k\in K^S}$ be the computing capacity allocation vector, so that the computing delay of user $u_k$ processing its computing task on SBS $k$ is denoted as
\begin{equation}
  \label{deqn_cd2}
  T^O_{u_k} = \frac{\sum \limits ^M_{m=1}z_{u_km}F_{u_km}\left(b_{u_k}\right)}{f^O_{u_k}},\forall u,k.
\end{equation}
In our model, we need to consider the influence of multi-task parallel computing of MEC severs. According to the virtual machine multiplexing technology, we adopt the influence of multi-task parallel computing on delay of computing of computing task $m$ from \cite{ref3}, which is given as  
\begin{equation}
  \label{deqn_pd}
  T_{u_km} = T^O_{u_k} \left(1+d\right)^{i_m-1},\forall u,k,m,
\end{equation}
where $i_m$ represents the amount of parallel computing requested when classification intelligent task $m$ is processed, and $d(\geq 0)$ is degradation factor which means the percentage increase in the expected computing delay experienced by a virtual machine when multiplexed with another virtual machine. 
Therefore, $d$ is used to represent the impact of multi-task parallel computing on MEC servers. We assume that a maximum of $Q$ parallel numbers are allowed to be processed simultaneously on the MEC server, i.e., $i_m \leq Q$. Therefore, the delay of computing of user $u_k$ on SBS $k$ is denoted as
\begin{equation}
  \label{deqn_doS}
  t^{comp}_{u_k} = \sum^M_{m=1} z_{u_km}T_{u_km}, \forall u,k.
\end{equation}

Similar to the study in \cite{ref19}, we notice the fact that the downlink data volume of computing outcome is much smaller than uplink data volume, so we neglect the downlink transmission in this work. Due to the character of local computing, we do not consider the influence of multi-task parallel computing on local device. 

\subsection{Utility Function}
As shown from above introduction, we consider joint allocation of communication resource and computing capacity with compression and parallel computing. In this paper, we focus on maximum the system utility under computing accuracy constraint and task delay constraint. For each user $u_k$, we consider marginal utility of the combination of system revenue, i.e., computing accuracy and system cost, i.e., task delay. 

To design intelligent computing and introduce the feature of classification intelligence tasks in our model, we adopt the 3-parameters power law fitting formula between the data volume and the computing accuracy from \cite{ref2}. Note that the accuracy fitting formula of classification intelligence tasks is used to represent the training process which is different from the inference accuracy in \cite{ref14,ref30}. Obviously, we consider the scenario of distributed few-shot learning and design the resource allocation in this paper, and the inference accuracy will be considered in our future work. 
For the convenience of subsequent modelling, we adopt the simplified form which is given as 
\begin{equation}
  \label{deqn_fit}
  y(\alpha_{u_k}) = p - q \alpha_{u_k}^{-r},\forall u,k,
\end{equation}
where $\alpha_{u_k}$ represents the data volume to be computed of user $u_k$ and $p,q,r$ are all fitting paraments.
In this work, the limit of computing accuracy of computing task $m$ is set as $\tilde{y}_m$. Therefore, the computing accuracy of user $u_k$ in our model is denoted as
\begin{equation}
  \label{deqn_acc}
  y(\alpha_{u_k}) = p - q ((1-x_{u_k})a_{u_k} + x_{u_k}b_{u_k})^{-r}.
\end{equation}
In scenario of classification intelligence tasks, the raw data $a_{u_k}$ is training samples for several mode and we consider the raw data as data volume for resource allocation. 
Note that we adopt $b_{u_k}$ as the data volume for accuracy calculation after compression and transmission. This mode is used for the lite distributed training system that need to make trade-off between delay and accuracy.
The total task delay of user $u_k$ in our model is
\begin{equation}
  \label{deqn_tod}
  t_{u_k} = (1-x_{u_k})T^L_{u_k} + x_{u_k}(t^{comm}_{u_k}+t^{comp}_{u_k}), \forall u,k.
\end{equation}
Note that we consider the two different properties, i.e., task delay and computing accuracy, of classification intelligent tasks which is different from traditional computing tasks. In the system, these two properties are the primary concern in practice \cite{ref31}.
To moderate the trend of accuracy and delay and balance the complexity of algorithm, we model the system utility with convex and nondecreasing function. Here the logarithmic function of diminishing marginal utility with trade-off between system revenue and cost which has been used frequently in literature\cite{ref4}, is adopted as utility function. Therefore the utility of user $u_k$ is denoted as
\begin{equation}
  \label{deqn_uu}
  R_{u_k} = ln\left( L \frac{y(\alpha_{u_k})}{t_{u_k}} \right), \forall u,k,
\end{equation}
where $L$ is denoted as the weight parameter between system revenue and cost. We adopt the form of division and weight parameter in logarithmic function to control the order of magnitude of delay because the change in value of accuracy is small. 
Obviously, the equation \eqref{deqn_uu} represents the marginal utility of users while being processed in this system considering trade-off between accuracy and delay.
Therefore, the system utility is given as
\begin{equation}
  \label{deqn_su}
  R = \sum_{k\in K^S}\sum_{u_k\in U^S_k} R_{u_k},
\end{equation}
which is the system optimization objective considered in this paper. 

\section{Problem Formulation and Algorithm Design}
In order to maximize the system utility, we formulate it as an optimization problem and decompose it into several convex optimization subproblems via successive convex approximation (SCA). 
Then we design the corresponding iterative algorithm to solve the optimization problem.

\subsection{Problem Formulation and Decomposition Solution}
We adopt the system utility proposed in \eqref{deqn_su} as the objective function of our optimization problem, and we formulate it as
\begin{equation}
  \begin{aligned}
  \label{opt}
  &\ \ \, \max_{\bm{x},\bm{\rho},\bm{f^O},\bm{\varepsilon}} R  \\ 
  &{\rm{s.t.}}\ (\mathrm{C}1): x_{u_k}\in \{0,1\},  \forall u,k,  \\
  &\ \ \ \ \ (\mathrm{C}2): \rho_{u_kn} \in \{0,1\}, \forall u,k,n, \\
  &\ \ \ \ \ (\mathrm{C}3):\sum^K_{k=1} x_{u_k} \leq 1, \forall u, \\
  &\ \ \ \ \ (\mathrm{C}4):\sum^{U_k}_{u_k=1} \rho_{u_kn} \leq 1, \forall n,\\
  &\ \ \ \ \ (\mathrm{C}5):\varepsilon_{u_k} \geq 1, \forall u,k, \\
  &\ \ \ \ \ (\mathrm{C}6):t_{u_k} \leq \sum^M_{m=1} z_{u_km}\tilde{t}_m, \forall u,k,  \\
  &\ \ \ \ \ (\mathrm{C}7):y(\alpha_{u_k})\geq \sum^M_{m=1} z_{u_km}\tilde{y}_m, \forall u,k,  \\
  &\ \ \ \ \ (\mathrm{C}8):\sum^{U_k}_{u_k =1}f^O_{u_k} \leq F_k, \forall u,k.
  \end{aligned}
\end{equation}
In \eqref{opt}, the constraints (C1) and (C2) guarantees that the value of two indicator variables is restrict to 0 and 1, 
constraints (C3), (C4) and (C5) mean one user can only choose one type of computing approach, one subcarrier $n$ on one SBS $k$ can only be allocated to one user $u_k$ at a time and the compressed data is less than or equal to the raw data, 
constraints (C6) and (C7) are proposed to ensure the limits of the task delay and the computing accuracy are hold,  
constraint (C8) guarantees that the sum of allocated computing capacity is not greater than total computing capacity of MEC servers.

It is observed that constraints (C5) and (C7) is the external to the system which is chosen by users, which means compression ratio of terminals affects the resource allocation of system to large extent and the computing accuracy constraint can change the traditional resource allocation which will be displayed by the benchmark in simulation below. Obviously, the dimensions of variable $\bm{\rho}$ is the most and will pull the computing complexity of the algorithm below deeply.

Apparently, \eqref{opt} is a non-linear mixed integer programming and non-convex optimization problem, and such problems are usually considered as NP-hard problems. Therefore, we need to decompose it into several subproblems and make some transformation and simplification to solve it iteratively. 
For convenience of solving \eqref{opt}, we decompose it into three subproblems by giving other variables. 

\emph{1) Subcarrier Allocation Subproblem:} In this subproblem, other variables are give in \eqref{opt} except subcarrier allocation variable. In order to facilitate the analysis and solution of the problem, the method of binary variable relaxation \cite{ref20} is adopted, variable $\bm{\rho}$ is relaxed into real value variable as $\rho_{u_kn} \in [0,1]$. Then we can acquire subcarrier allocation subproblem which is given as 
\begin{equation}
  \begin{aligned}
  \label{sub_rho}
  &\ \ \ \ \, \max_{\bm{\rho}} \sum_{k\in K^S}\sum_{u_k\in U^S_k}ln\left( \frac{L A^\delta_{u_k}}{A^\beta _{u_k}+x_{u_k}\left( \frac{b_{u_k}}{r_{u_k}} + A^\gamma_{u_K} \right) } \right) \\
  &{\rm{s.t.}} \ (\mathrm{C}2), (\mathrm{C}4), \\
  &\ \ \ \ \ (\mathrm{C}6'): A^\beta _{u_k}+x_{u_k}\left( \frac{b_{u_k}}{r_{u_k}} + A^\gamma_{u_K} \right)\leq \sum^M_{m=1} z_{u_km}\tilde{t}_m, \forall u,k,
  \end{aligned}
\end{equation}
where $A^\delta_{u_k} = p - q ((1-x_{u_k})a_{u_k} + x_{u_k}b_{u_k})^{-r}$, $A^\beta _{u_k} = (1-x_{u_k})T^L_{u_k}$ and $A^\gamma_{u_K} = t^{comp}_{u_k}$ and in this way, (C6) in \eqref{opt} is converted to $(\mathrm{C}6')$ here. These expressions do not vary with $\bm{\rho}$, therefore we use constant expressions to replace them for convenience of solving \eqref{sub_rho}. $r_{u_k}$ is expanded out to 
\begin{equation}
  \begin{aligned}
  \label{deqn_r}
  r_{u_k} = &\sum^N_{n=1} \rho_{u_kn} \frac{B}{N} \cdot \\
  & log_2\left( 1 + \frac{P_{u_kn}g_{u_kn}}{\sum \limits_{c\in K^S,c\neq k} \sum \limits^{U_c}_{u'_c=1} \rho_{u'_cn}g_{u'_cn}P_{u'_cn} + \sigma^2} \right).
  \end{aligned}
\end{equation}
Obviously, the relationship between $\bm{\rho}$ and the objective function of \eqref{sub_rho} is pretty complicated. For convenience of expressing the solution of this problem, we remove constant terms and replace the interference item with a single variable, and get the optimization problem which is denoted as
\begin{equation}
  \begin{aligned}
  \label{sub_transrho}
  &\ \ \ \ \, \min_{\bm{\rho},\bm{I}} \sum_{k\in K^S}\sum_{u_k\in U^S_k}ln\left( \frac{x_{u_k}b_{u_k}}{\frac{B}{N}\sum \limits ^N_{n=1} log_2\left( 1 + \frac{P_{u_kn}g_{u_kn}}{I_{u_kn} + \sigma^2} \right)}\right) \\
  &{\rm{s.t.}} \ (\mathrm{C}2), (\mathrm{C}4), \\
  &\ \ \ \ \ (\mathrm{C}6'): \frac{x_{u_k}b_{u_k}}{\frac{B}{N}\sum \limits ^N_{n=1} log_2\left( 1 + \frac{P_{u_kn}g_{u_kn}}{I_{u_kn} + \sigma^2} \right)} \leq \\
  &\qquad \qquad \ \ \sum^M_{m=1} z_{u_km}\tilde{t}_m - A^\beta _{u_k} - x_{u_k}A^\gamma_{u_K}, \forall u,k,   \\
  &\ \ \ \ \ (\mathrm{C}9): I_{u_kn} = \sum_{c\in K^S,c\neq k} \sum^{U_c}_{u'_c=1} \rho_{u'_cn}g_{u'_cn}P_{u'_cn},\forall u,k,n, 
  \end{aligned}
\end{equation}
where $\bm{I} = \{ I_{u_kn}\}_{u_k\in U^S_k, k\in K^S, n\in N^S}$ represent the interference vector and become constraint (C9) of the above problem. Apparently, \eqref{sub_transrho} is non-convex problem, therefore we adopt the method of SCA to transform it and relaxation variables are introduced as follow
\begin{equation}
  \begin{aligned}
    \label{deqn_rel}
    &S_{u_kn} \leq \rho_{u_kn}l_{u_kn},\\
    &l_{u_kn} \leq log_2 \left( 1 + \frac{P_{u_kn}g_{u_kn}}{I_{u_kn} + \sigma^2}  \right),
  \end{aligned}
\end{equation}
which will be constraints $(\mathrm{C}7')$ and $(\mathrm{C}8')$ of the problem. Then the original optimization problem \eqref{sub_transrho} is converted to
\begin{equation}
  \begin{aligned}
  \label{sub_relaxsub}
  &\ \ \ \, \min_{\bm{\rho},\bm{I},\bm{S},\bm{l}} \sum_{k\in K^S}\sum_{u_k\in U^S_k}ln\left( \frac{x_{u_k}b_{u_k}}{\frac{B}{N}\sum \limits ^N_{n=1}S_{u_kn} }\right) \\
  &{\rm{s.t.}} \ (\mathrm{C}2), (\mathrm{C}4),(\mathrm{C}6'),\\
  &\ \ \ \ \ (\mathrm{C}7'): S_{u_kn} \leq \rho_{u_kn}l_{u_kn},\\
  &\ \ \ \ \ (\mathrm{C}8'): l_{u_kn} \leq log_2 \left( 1 + \frac{P_{u_kn}g_{u_kn}}{I_{u_kn} + \sigma^2}  \right),\\
  &\ \ \ \ \ (\mathrm{C}9): I_{u_kn} = \sum_{c\in K^S,c\neq k} \sum^{U_c}_{u'_c=1} \rho_{u'_cn}g_{u'_cn}P_{u'_cn},\forall u,k,n.
  \end{aligned}
\end{equation}
It shows that the right side of $(\mathrm{C}7')$ and $(\mathrm{C}8')$ require SCA to convert them to convex, and others constraints are all convex. 
For $(\mathrm{C}7')$, we perform first order Taylor expansion on the right side at point $(\rho^i_{u_kn},l^i_{u_kn})$ and convert it to
\begin{equation}
  \begin{aligned}
  \label{deqn_taex}
  S_{u_kn} \leq & \frac{\rho^i_{u_kn} + l^i_{u_kn}}{2} (\rho_{u_kn}+l_{u_kn}) - \\
  &\frac{\left(\rho^i_{u_kn} + l^i_{u_kn}\right)^2}{4} - \frac{\left(\rho^i_{u_kn} - l^i_{u_kn}\right)^2}{4},\forall u,k,n.
  \end{aligned}
\end{equation}
For $(\mathrm{C}8')$, we perform first order Taylor expansion on the right side at point $I^i_{u_kn}$ and convert it to
\begin{equation}
  \begin{aligned}
   \label{deqn_taex2}
    l_{u_kn} \leq &log_2(P_{u_kn}g_{u_kn} + I_{u_kn} + \sigma^2 ) - \\
    &\left( ln(I^i_{u_kn} + \sigma^2)+\frac{I_{u_kn}-I^i_{u_kn}}{I^i_{u_kn}+\sigma^2}  \right)/ln2, \forall u,k,n.
  \end{aligned}
\end{equation}
In this way $(\mathrm{C}7')$ and $(\mathrm{C}8')$ are converted to convex constraints and we can use convex optimization method for SCA iteration \cite{ref28} to solve \eqref{sub_relaxsub}, which is the same way to solve \eqref{sub_rho}.

\emph{2) Computing Capacity Allocation Subproblem:} Under given other variables except $\bm{f^O}$, \eqref{opt} is simplified to
\begin{equation}
  \begin{aligned}
    \label{sub_comp}
    &\ \ \ \ \ \max_{\bm{f^O}} \sum_{k\in K^S}\sum_{u_k\in U^S_k}ln(LA^\delta _{u_k}) - ln (B^\beta _{u_k}+t^{comp}_{u_k})\\
    &{\rm{s.t.}} \ (\mathrm{C}6''): B^\beta _{u_k}+t^{comp}_{u_k} \leq \sum^M_{m=1} z_{u_km}\tilde{t}_m,\forall u,k,\\
    &\ \ \ \ \ (\mathrm{C}8),
  \end{aligned}
\end{equation}
where the constant term $B^\beta _{u_k} = (1-x_{u_k})T^L_{u_k} + x_{u_k}t^{comm}_{u_k}$, and $t^{comp}_{u_k} = \sum^M_{m=1} z_{u_km} (1+d)^{i_m-1} \cdot \\ \frac{\sum^M_{m=1}z_{u_km}F_{u_km}(b_{u_k})}{f^O_{u_k}}$ and in this way, (C6) in \eqref{opt} is converted to $(\mathrm{C}6'')$ here. Therefore, \eqref{sub_comp} is a convex optimization problem and can be solved directly by convex optimization method \cite{ref21}.

\emph{3) Compression Offloading Subproblem:} We need to solve computing offloading indicator variable $\bm{x}$ and compression ratio variable $\bm{\varepsilon}$ under given $\bm{\rho}$ and $\bm{f^O}$. For convenience of solving, we adopt binary variable relaxation and relax $\bm{x}$ into real variable as $x_{u_k} \in \{ 0,1 \}$.
The original problem \eqref{opt} is simplified to
\begin{equation}
  \begin{aligned}
    \label{sub_comoff}
    &\ \ \ \ \, \max_{\bm{x},\bm{\varepsilon}} \sum_{k\in K^S}\sum_{u_k\in U^S_k} ln\left( \frac{L y(\alpha_{u_k})}{(1-x_{u_k})C^\delta_{u_k}+\frac{x_{u_k}}{\varepsilon_{u_k}}C^\beta_{u_k}} \right)\\
    &{\rm{s.t.}} \ (\mathrm{C}1), (\mathrm{C}3),(\mathrm{C}5),\\
    &\ \ \ \ \ (\mathrm{C}6'''): (1-x_{u_k})C^\delta_{u_k}+\frac{x_{u_k}}{\varepsilon_{u_k}}C^\beta_{u_k} \leq \sum^M_{m=1} z_{u_km}\tilde{t}_m, \forall u,k,\\
    &\ \ \ \ \ (\mathrm{C}7),
  \end{aligned}
\end{equation}
where the constant terms $C^\delta_{u_k} = T^L_{u_k}$ and $C^\beta_{u_k} = \frac{a_{u_k}}{r_{u_k}} + \sum^M_{m=1}z_{u_km}(1+d)^{i_m-1}\frac{\sum^M_{m=1}z_{u_km}F_{u_km}(a_{u_k})}{f^O_{u_k}}$, and $y(\alpha_{u_k}) = p - q ((1-x_{u_k})a_{u_k} + \frac{x_{u_k}}{\varepsilon_{u_k}}a_{u_k})^{-r}$ and in this way, (C6) in \eqref{opt} is converted to $(\mathrm{C}6''')$ here.
Normally, $p,q$ and $r$ satisfy that $p>0, q>0$ and $0\leq r \leq 1$. We adopt the method of variable substitution and let $\eta_{u_k} = 1-x_{u_k}+\frac{x_{u_k}}{\varepsilon_{u_k}}$. Obviously, $\eta_{u_k}$ satisfies that $1-x_{u_k}\leq\eta_{u_k}\leq1$ which will be constraint $(\mathrm{C}5')$ of the above problem and we can transform problem \eqref{sub_comoff} into 
  \begin{align}
    \label{sub_trans_comoff}
    &\max_{\bm{x},\bm{\eta}} \sum_{k\in K^S}\sum_{u_k\in U^S_k} ln\left( \frac{L (p-q*(a_{u_k}\eta_{u_k})^{-r})}{(1-x_{u_k})(C^\delta_{u_k}-C^\beta_{u_k})+C^\beta_{u_k}\eta_{u_k}} \right) \notag \\
    &s.t. \ (\mathrm{C}1),(\mathrm{C}3), \notag \\ 
    &\ \ \ \ \ (\mathrm{C}5'): 1-x_{u_k} \leq \eta_{u_k}\leq 1, \forall u,k,\notag \\ 
    &\ \ \ \ \ (\mathrm{C}6'''): (1-x_{u_k})(C^\delta_{u_k}-C^\beta_{u_k})+C^\beta_{u_k}\eta_{u_k} \leq \\
    &\qquad \qquad \qquad \qquad \qquad \qquad \quad \ \ \sum^M_{m=1} z_{u_km}\tilde{t}_m, \forall u,k,\notag \\
    &\ \ \ \ \ (\mathrm{C}7): p-q*(a_{u_k}\eta_{u_k})^{-r} \geq \sum^M_{m=1} z_{u_km}\tilde{y}_m,\forall u,k.\notag
  \end{align}
Due to non-convexity of \eqref{sub_trans_comoff}, we adopt the method of SCA and let
\begin{equation}
  \label{deqn_v}
  v_{u_k} \geq ln\left( (1-x_{u_k})(C^\delta_{u_k}-C^\beta_{u_k}) + C^\beta_{u_k}\eta_{u_k} \right).
\end{equation}
We perform first order Taylor expansion on the right side at point $(x^j_{u_k},\eta^j_{u_k})$ and convert it to
\begin{equation}
  \begin{aligned}
  \label{deqn_v2}
  v_{u_k} \geq &ln\left( (1-x^j_{u_k})(C^\delta_{u_k}-C^\beta_{u_k}) + C^\beta_{u_k}\eta^j_{u_k} \right)+\\
  &\frac{(C^\beta_{u_k}-C^\delta_{u_k})(x_{u_k}-x^j_{u_k}) + C^\beta_{u_k}(\eta_{u_k} - \eta^j_{u_k})}{(1-x^j_{u_k})(C^\delta_{u_k}-C^\beta_{u_k}) + C^\beta_{u_k}\eta^j_{u_k}},
\end{aligned}
\end{equation}
which will be constraint (C10) of the above problem. Therefore, \eqref{sub_trans_comoff} is converted to
\begin{equation}
  \begin{aligned}
    \label{sub_sca_comoff}
    &\max_{\bm{x},\bm{\eta}} \sum_{k\in K^S}\sum_{u_k\in U^S_k} ln\left( \frac{L(p-q*(a_{u_k}\eta_{u_k})^{-r})}{v_{u_k}} \right)\\
    &s.t. \ (\mathrm{C}1),(\mathrm{C}3), (\mathrm{C}5'),(\mathrm{C}6'''),(\mathrm{C}7),\\
    &\ \ \ \ \ (\mathrm{C}10): \eqref{deqn_v2}.
  \end{aligned}
\end{equation}
Then we can use convex optimization method for SCA iteration to solve \eqref{sub_trans_comoff} by using standard CVX tools\cite{ref29}.

\begin{algorithm}[H]
  \caption{Computing Offloading and Resource Allocation Algorithm for Intelligent Tasks}\label{alg:alg1}
  \begin{algorithmic}[1] 
   \STATE Set initial $q=0$, computing offloading variable $x^q$, computing capacity allocation $f^q$ and compression ratio $\varepsilon^q$.
   \STATE Set the iteration constraint $\theta > 0$.
   \REPEAT
   \STATE $q = q + 1$.
   \STATE Obtain $\rho^q$ by solving subcarrier allocation subproblem \eqref{sub_rho} through $x^{q-1}$, $f^{q-1}$ and $\varepsilon^{q-1}$:
   \STATE Set initial $i=0$, the SCA iteration constraint $\theta_1 > 0$, $\rho^i_{u_kn}$ and $l^i_{u_kn}$ according to \eqref{deqn_taex} and \eqref{deqn_taex2}.
   \REPEAT
   \STATE $i = i + 1$.
   \STATE Obtain $\rho^i_{u_kn}$ and $l^i_{u_kn}$ by solving convex optimization problem \eqref{sub_relaxsub}.
   \STATE Obtain the value of \eqref{sub_transrho}, i.e., $N^i_{sub1}$.
   \UNTIL $\left\lvert N^i_{sub1}-N^{i-1}_{sub1}\right\rvert \leq \theta_1$.
   \STATE Obtain $f^q$ by solving computing capacity allocation subproblem \eqref{sub_comp} through $x^{q-1}(t)$, $\varepsilon^{q-1}$ and $\rho^{q}$ directly by convex optimization.
   \STATE Obtain $x^q$ and $\varepsilon^q$ by solving compression offloading subproblem \eqref{sub_comoff} through $\rho^q$ and $f^q$:
   \STATE Set initial $j=0$, the SCA iteration constraint $\theta_3 > 0$, $x^j_{u_k}$ and $\eta^j_{u_k}$ according to \eqref{deqn_v2}.
   \REPEAT
   \STATE $j = j + 1$.
   \STATE Obtain $x^j_{u_k}$ and $\eta^j_{u_k}$ by solving convex optimization problem \eqref{sub_sca_comoff}.
   \STATE Obtain the value of \eqref{sub_comoff}, i.e., $N^j_{sub3}$.
   \UNTIL $\left\lvert N^j_{sub3}-N^{j-1}_{sub3}\right\rvert \leq \theta_3$.
   \STATE Obtain the value of \eqref{opt}, i.e., $N^q$ through $x^q$, $\rho^q$, $f^q$ and $\varepsilon^q$
   \UNTIL $\left\lvert N^q-N^{q-1}\right\rvert \leq \theta$.
  \end{algorithmic}
\end{algorithm}

\subsection{Algorithm Design and Analysis}
As mentioned above, we decompose the original NP-hard problem \eqref{opt} into three subproblems. Then we use the idea of the greedy algorithm to iterate the above solutions of three subproblems and arrive at the suboptimal solution for \eqref{opt}, which is summarized in Algorithm 1.

In Algorithm 1, we adopt alternating iteration of three problems and obtain the solutions in closed forms by convex optimization. According to the greedy algorithm and convex optimization theory \cite{ref22}, iteration of three subproblems can ensure $\left\lvert N^q-N^{q-1}\right\rvert \leq \theta$, i.e., convergence quickly but only sub-optimality can be guaranteed \cite{ref23}. 
As we show above, the complexity of Algorithm 1 depends on three subproblems. In subproblem 1, since \eqref{sub_relaxsub} is solved through SCA iteration by converting constraints, we assume the number of iterations is $L_{sub1}$, then the complexity is $\mathcal{O}((UN)^2L_{sub1})$. In subproblem 2, since \eqref{sub_comp} is a convex optimization problem, the complexity is $\mathcal{O}(U)$. In subproblem 3, \eqref{sub_comoff} need to be converted to \eqref{sub_sca_comoff} through SCA and achieve solution in iteration algorithm, we assume the number of iterations is $L_{sub3}$, therefore the complexity is $\mathcal{O}(UL_{sub3})$. We assume the number of total iteration is $L_{it}$, then the overall complexity of Algorithm 1 is $\mathcal{O}(((UN)^2L_{sub1}+U+UL_{sub3})L_{it})$.
In the way, the NP-hard optimization problem \eqref{opt} is decomposed into low-complexity subproblems and iteratively solved.

\section{Simulation Result}
In this section, we first set the simulation paraments and then show our simulation results to evaluate the performance of our proposed algorithm.

\subsection{Simulation Parameters}
We consider system level simulation of uplink transmission in a small cell heterogeneous cellular scenario according to the 3GPP normative document of small cell network, i.e., urban micro (UMi) model \cite{ref24,ref25}, which is considered to be deployed in separate channels from MBS and we adopt the hexagonal cell deployment mode. 
The distance between the user and SBS meet the standard of the 3GPP normative document which indicates that there is no interference between SBS and MBS, and only outdoor links exist. 
In our model, we consider that four SBSs are deployed in a small cell area with a total coverage of $200{\rm{m}} \times 200{\rm{m}}$. The SBSs provide offloading association and resource allocation for users. Suppose that there are LoS and NLoS links in our scenario. Let $d_{u_k}$ be the distance between SBS $k$ and user $u_k$, and note that the path loss depend on the link state of LoS and NLoS \cite{ref25}. Therefore, when it is a LoS link, the path loss of user $u_k$ is given by
\begin{equation}
  \label{deqn_pllos}
  \mu^{LoS}_{u_k} = 22.0log_{10}(d_{u_k}) + 28.0 + 20log_{10}(F^q),
\end{equation}
and when it is a NLoS link, the path loss is given by 
\begin{equation}
  \label{deqn_plnlos}
  \mu^{NLoS}_{u_k} = 36.7log_{10}(d_{u_k}) + 22.7 + 26log_{10}(F^q),
\end{equation}
where $F^q$ is the carrier frequency. The LoS probability that determines the link state is denoted as 
\begin{equation}
  \label{deqn_pro}
  p^{LoS}_{u_k} = \min \left( \frac{18}{d_{u_k},1} \right) \left( 1 - e^{-\frac{d_{u_k}}{36}} \right) + e^{-\frac{d_{u_k}}{36}},   
\end{equation}
and the NLoS probability is $p^{NLoS}_{u_k} = 1 -p^{LoS}_{u_k}$. Therefore, according to \cite{ref24,ref25} the channel gain in small cell network in this scenario is denoted as
\begin{equation}
  \label{deqn_chag}
  g_{u_k} = \left( p^{LoS}_{u_k}10^{\mu^{LoS}_{u_k}} + p^{NLoS}_{u_k}10^{\mu^{NLoS}_{u_k}} \right) ^{-1}.
\end{equation}

In our system model, we consider two processes including computing accuracy and parallel computing for computing tasks. The paraments of them are set according to the most suitable fitting paraments \cite{ref2,ref3}. 
Simulation paraments are summarized in Tabel II\cite{ref25}.
\begin{table}[h]
  \renewcommand\arraystretch{1.5}
  \caption{Simulation Parameters}
  \centering
  \begin{tabular}{p{4.8cm}<{\centering}|p{2.1cm}<{\centering}}
  \hline
  \hline
  {\textbf{Parameter}} & {\textbf{Value}} \\
  \hline
  Bandwidth resource, $W_k$ & 10 MHz \\
  \hline
  Transmit power, $P_{u_kn}$ & 0.1 W \\
  \hline
  The noise power, $\sigma^2$ & -100 dBm \\
  \hline
  The carrier frequency, $F^q$ & 3.5 GHz\\
  \hline
  The number of subcarriers, $N$ & 64 \\
  \hline
  Computing capacity of MEC servers, $F_k$ & 200 Gigacycle/s \\
  \hline
  Computing capacity of local device, $F^L_{u_k}$ & 1.4 Gigacycle/s \\
  \hline
  The total number of users, $U$ & 30\\
  \hline
  The weight parameter, $L$ & 1\\
  \hline
  The number of iterations, $L_{it}$ & 10\\
  \hline
  Fitting paraments of computing accuracy, $p,q,r$ & 100, 80, 0.6\\
  \hline
  Maximum number of parallel tasks, $Q$ & 5\\
  \hline
  Degradation factor, $d$ & 0.2\\
  \hline 
  \hline
  \end{tabular}
\end{table}

According to the computing delay and accuracy requirements of some services of ultra reliable low latency communications \cite{ref26,ref27}, we assume there are three task types in our simulation and the requirements are different. The delay and accuracy limits of tasks are shown in Tabel III.
\begin{table}[h]
  \renewcommand\arraystretch{1.5}
  \caption{Task Parameters}
  \centering
  \begin{tabular}{p{1.5cm}<{\centering}|p{1.8cm}<{\centering}|p{3.1cm}<{\centering}}
    \hline
    \hline
    {\textbf{Task type}} & {\textbf{Task delay}} \bm{$\tilde{t}_m$} & {\textbf{Computing accuracy}} \bm{$\tilde{y}_m$}\\
    \hline
    1 & 20 ms & 85 \\ 
    \hline
    2 & 40 ms & 90 \\
    \hline
    3 & 60 ms & 95 \\
    \hline
    \hline
  \end{tabular}
\end{table}

\begin{figure}[!t]
  \centering
  \includegraphics[scale = 0.6]{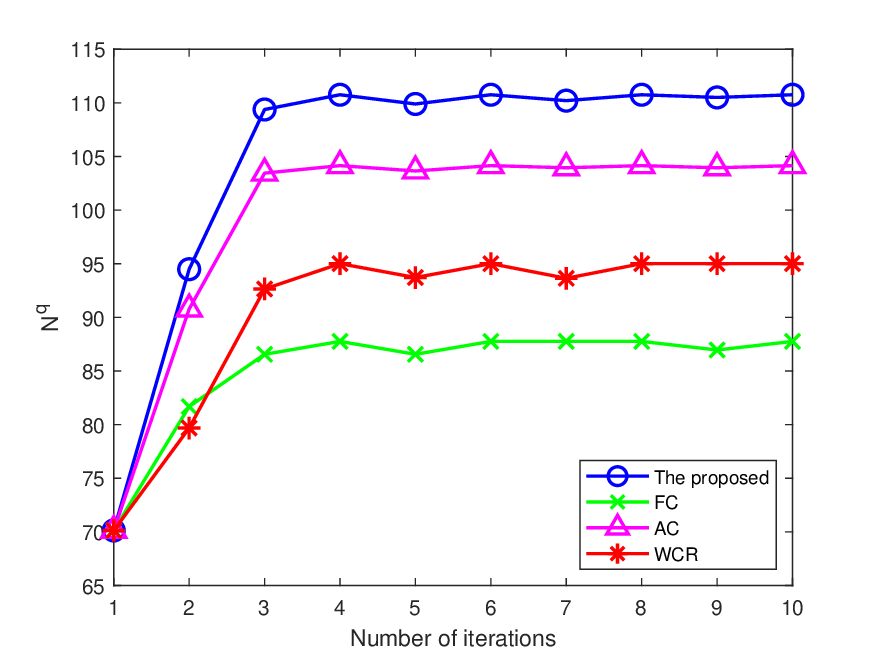}
  \caption{Convergence of all algorithms.}
  \label{fig_2}
\end{figure}

\subsection{Performance of the Proposed Algorithm}
In order to verify the performance of the proposed algorithm, we add the following schemes for comparison:
\begin{itemize}
\item{Fixed Channel (FC): The scheme is that subcarriers are allocated fixed and bandwidth is allocated evenly.}
\item{Average Computing (AC): The scheme is that computing capacity of MEC servers is allocate averagely.}
\item{Without Compression Ratio (WCR)\cite{ref10}: According to the scheme in \cite{ref10}, the compression ratio and parallel computing is not considered and computing offloading is processed directly.}
\end{itemize}

We demonstrate the convergence of all schemes in Fig. 2 and we can see that the convergence of our proposed algorithm is fast in $L_{it}$ iterations and the trend is basically fixed after convergence, which means our algorithm based SCA and iteration have a good stability and the astringency. From the convergence of comparison algorithm, we find that our proposed algorithm can acquire a better value of system utility and better optimization character in our system model considering joint allocation of communication resource and computing capacity. 

\begin{figure}[!t]
  \centering
  \includegraphics[scale = 0.6]{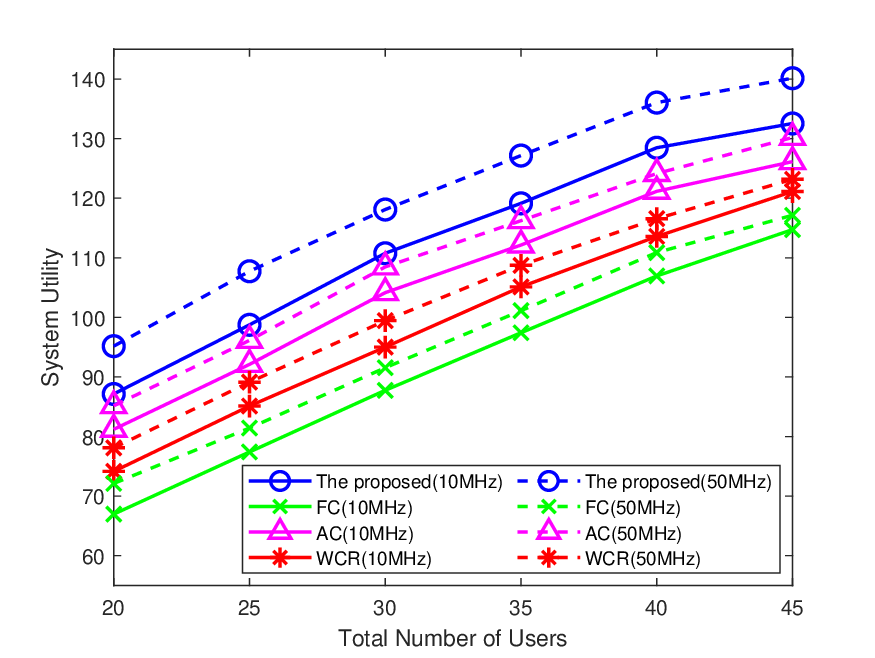}
  \caption{System utility varying with total number of users under different bandwidth.}
  \label{fig_3}
\end{figure}

The characteristics of system utility with total number of users $U$ under different bandwidth, i.e., 10 MHz and 50 MHz, as Fig. 3 shows.
It is found that the system utility increases with the total number of users and the trend is slower when total number of users is greater than 35 in our proposed algorithm.
When total number of users is relatively small, the resources are sufficient and resource allocation is efficient, therefore the system utility increases quickly. Nevertheless, as total number of users is relatively large, the resources of system is limited and resource allocation will become inefficient, the growth tendency of system utility will slow down.
The comparison schemes all have this property but the trend is not notable, which is different for different algorithms. For example, the trend of FC is the least significant because the fixed allocation of communication resource result in inefficient resource allocation and is less affected by resource limits.
We can see in this figure that the higher bandwidth has larger impact in our proposed algorithm than other comparison schemes which means our scheme have higher usage in bandwidth. 
\begin{figure}[!t]
  \centering
  \includegraphics[scale = 0.6]{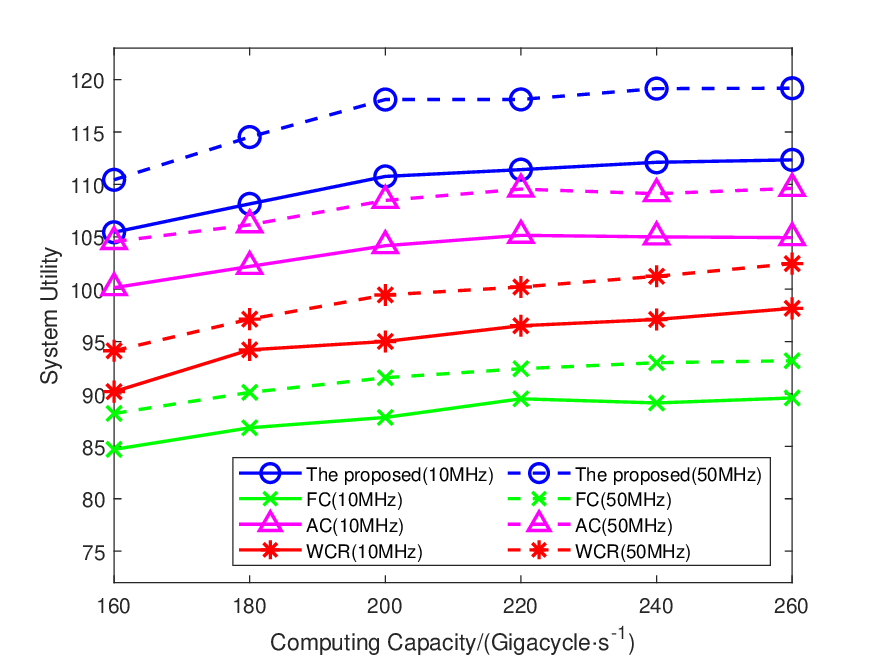}
  \caption{System utility varying with computing capacity of MEC servers under different bandwidth.}
  \label{fig_4}
\end{figure}

We compare system utility with computing capacity $F_k$ of MEC servers under different bandwidth in Fig. 4. From the trend we can see that there is a inflection point of system utility in $F_k = 200$ Gigacycle/s in our proposed algorithm. 
This is because we consider the computing accuracy limit in our system model, the system utility depends on computing accuracy and task delay and our proposed algorithm need make a trade-off between them. We can get a better trade-off when $F_k$ is relatively small and reaches a certain value while increasing. However, the communication resource will be limited and affects the compression ration and limits computing accuracy when $F_k$ continues to rise, therefore users would choose local computing which result in a slowdown in growth of the system utility.
This property also presents in comparison algorithms FC and AC with different inflection points, but in WCR where compression ratio is not considered, the trade-off does not existed while $F_k$ is increasing. 
Also, we can see that higher bandwidth do not have a significant impact on this trend of system utility.
\begin{figure}[!t]
  \centering
  \includegraphics[scale = 0.6]{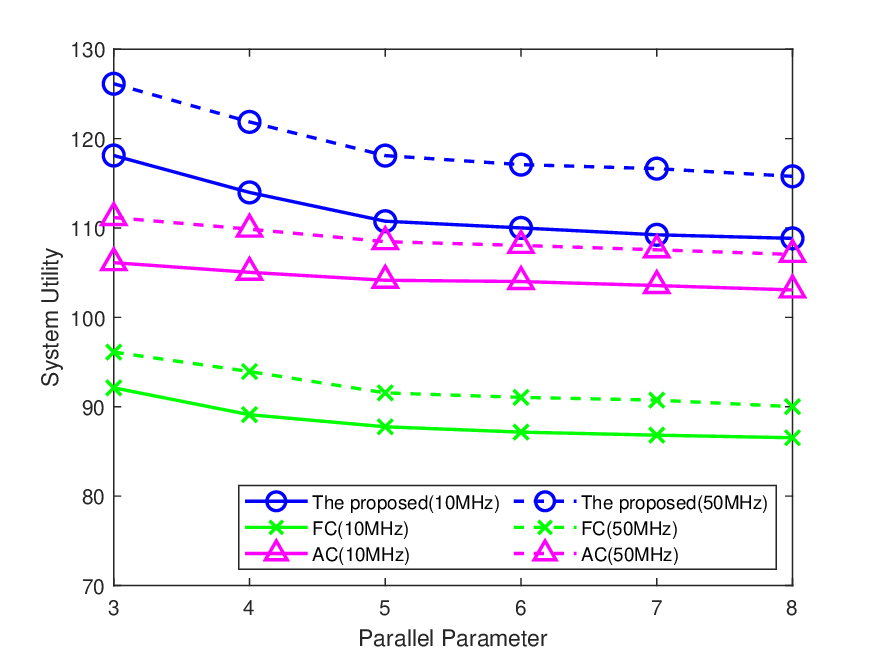}
  \caption{System utility varying with parallel parameter under different bandwidth.}
  \label{fig_5}
\end{figure}

The trend of system utility varying with parallel parameter $Q$ under different bandwidth is represented in Fig. 5 which is used to evaluate the property of parallel computing of classification intelligent tasks.
We can find that the system utility decreases as the $Q$ increases in our proposed algorithm, which means parallel computing has a important impact on our system model. 
In our system, The increase in $Q$ means that classification intelligent tasks will request more parallelism, which will cause that the computing delay increases and have a impact on total offloading process. Therefore, the system utility would decrease because the trade-off between computing accuracy and task delay is affected. 
This property also presents in comparison algorithms but it is not that significant in AC, which is because the average allocation of computing capacity would reduce the impact of $Q$. The trend of WCR is not represented because parallel computing is not considered. 
Moreover, the higher bandwidth have some influence on this trend but not notable in our system model.  
\begin{figure}[!t]
  \centering
  \includegraphics[scale = 0.6]{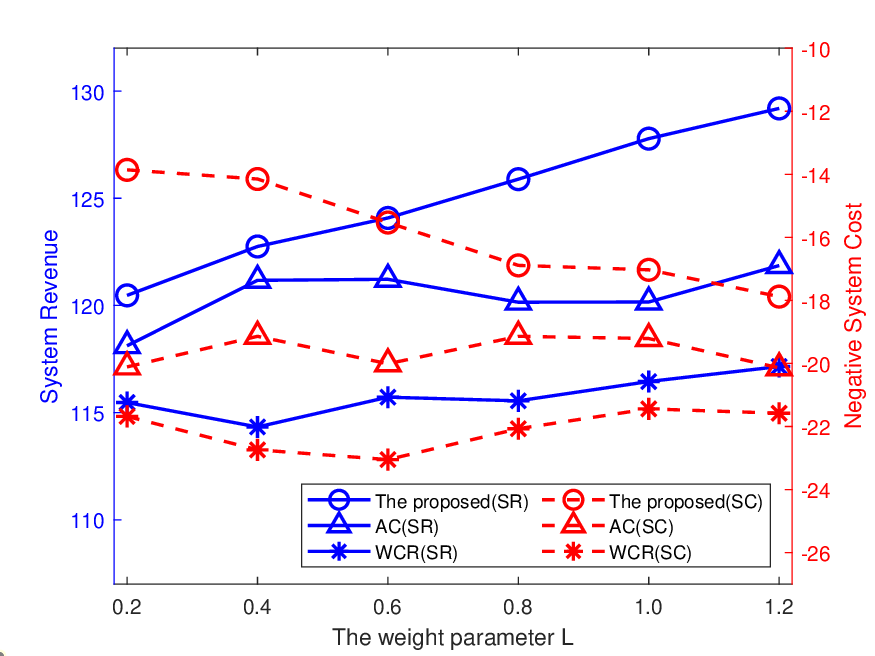}
  \caption{System revenue and negative system cost varying with the weight parameter.}
  \label{fig_6}
\end{figure}

We plot the trade-off between system revenue and negative system cost vs. $L$ in Fig. 6, where SR means system revenue and SC means negative system cost. 
Note that the negative system cost of our proposed algorithm is the largest which means system cost is the smallest, which make the representation of Fig. 6 more intuitional.
The system utility in \eqref{deqn_su} indicates that our proposed algorithm can balance system revenue, which is based on computing accuracy, and system cost, which is based on task delay. In this figure we can see that our proposed algorithm can get a better trade-off than comparison algorithm while $L$ increases. 
It is obvious that the decrease on system revenue and increase on system cost while $L$ is growing in our proposed algorithm, but this trend is not significant in other comparison schemes. Therefore, our proposed algorithm can balance revenue on computing accuracy and cost on task delay significantly to stabilize the increase of the system utility, and can acquire a better trade-off to make systems increasingly stable compared with comparison algorithms.

\section{Conclusion}
In this paper, we investigated computing offloading and resource allocation for intelligent tasks in MEC systems.
Specially, we focus on classification intelligence tasks and formulate an optimization problem that considers both the accuracy requirements of tasks and the parallel computing capabilities of MEC systems.
We decomposed it into three subproblems: subcarrier allocation, computing capacity allocation and compression offloading which were solved iteratively through convex optimization and successive convex approximation.
Based on the solutions, we design an efficient computing offloading and resource allocation algorithm for intelligent tasks in MEC systems.
We focus on the specific intelligent tasks but our method can apply to other applications.
Simulation results have demonstrated that our proposed algorithm significantly improves the performance of intelligent tasks in MEC systems and achieves a flexible trade-off between system revenue and cost considering intelligent tasks compared with the benchmarks.

\vfill

\end{document}